UNIÃO DE ENSINO DO SUDOESTE DO PARANÁ – UNISEP
FACULDADE EDUCACIONAL DE FRANCISCO BELTRÃO – FEFB
COORDENADORIA DE TRABALHO DE CONCLUSÃO DE CURSO
SISTEMAS DE INFORMAÇÃO

GILBERTO LUIS DE CONTO JUNIOR

# DIABETIC RETINOPATHY DETECTION BY RETINAL IMAGE RECOGNIZING

**SUMMARY**

## FIGURE LIST

## SOURCE CODE LIST

## TABLE LIST

## LISTA DE GRÁFICOS

## LISTA DE ABREVIAÇÕES

ACPM........................................................ *American College of Preventive Medicine*

ASRS........................................................... *American Society of Retina Specialists*

AAO…............................................................ *American Academy of Ophthalmology*

CBO.....................................................................*Conselho Brasileiro de Oftalmologia*

CNN......................................................................*Convolutional Neural Network*

EMD........................................................................*Edema Macular Diabético*

HOSP......................................................................*Hospital de Olhos de São Paulo*

NEI...........................................................................*National Eye Institute*

NIDDK………………National Institute of Diabetes and Digestive and Kidney Diseases

PDR........................................................................*Proliferative Diabetic Retinopathy*

OCT........................................................................*Tomografia de coerência óptica*

OMS........................................................................*Organização Mundial Da Saúde*

OpenCV...........................................................*Open Source Computer Vision Library*

SBD........................................................................*Sociedade Brasileira de Diabetes*

WHO.......................................................................*World Health Organization*

# RESUMO

DE CONTO, G. L. J. **DIABETIC RETINOPATHY DETECTION BY RETINAL IMAGE RECOGNIZING.** 2019. Francisco Beltrão. Trabalho de Conclusão de Curso de Sistemas de Informação da Faculdade Educacional de Francisco Beltrão – FEFB.

Many people are affected by diabetes around the world. This disease may have type 1 and 2. Diabetes brings with it several complications including diabetic retinopathy, which is a disease that if not treated correctly can lead to irreversible damage in the patient's vision. The earlier it is detected, the better the chances that the patient will not lose vision. Methods of automating manual procedures are currently in evidence and the diagnostic process for retinopathy is manual with the physician analyzing the patient's retina on the monitor. The practice of image recognition can aid this detection by recognizing Diabetic Retinopathy patterns and comparing it with the patient's retina in diagnosis. This method can also assist in the act of telemedicine, in which people without access to the exam can benefit from the diagnosis provided by the application. The application development took place through convolutional neural networks, which do digital image processing analyzing each image pixel. The use of VGG-16 as a pre-trained model to the application basis was very useful and the final model accuracy was 82%.

***OBSERVATION: This is a translated version of the original Portuguese (Detecção de retinopatia diabética por reconhecimento de imagem da retina) without the theory part, and may have translation errors. To get the original document (pt-BR) contact this email: gilbertodeconto@hotmail.com**

## 1.0 INTRODUCTION

The NIDDK (National Institute of Diabetes and Digestive and Kidney Diseases) (2016) conceptualizes diabetes as a disease that occurs when the individual's blood sugar rate is high, this occurs when pancreas does not produce insulin (type 1) or the insulin produced is not enough to the glycemic control (type 2). According to diabetes care article (2004), by 2030, it is estimated that the number of diabetics in the world should grow to approximately 366 million people.

Diabetic Retinopathy can develop in people with diabetes, according to SBEM (Brazilian Society of Endocrinology and Metabology) (2010), DR is a complication that occurs when excess glucose in the blood damages blood vessels within the retina. The disease is the leading cause of preventable blindness in working age people (20-70 years), being identified in one third of people with diabetes, associated with risk factors, such as complications in the vascular system. (CHEUNG; MITCHELL; WONG, 2010).

The detection of this disease is currently done manually through an eye background examination (Fundus Exam) which the ophthalmologist analyzes the patient's retina and checks for possible anomalies that could exist manually (ASRS – American Society of Retina Specialists, 2016).

The purpose of this work is to automate the process of detecting and evaluating diabetic retinopathy, and for this, a system was built using artificial intelligence. The system uses image recognition to perform this evaluation, fed by retinal imaging, identifies whether or not the patient has the disease in the fourth stage (Proliferative Diabetic Retinopathy).

The work proposes the development of the application in Python, using the OpenCV library, which is used for Image Processing and Video I/O, Data Structure, Linear Algebra, Basic Graphical User Interface, and mainly image processing.

## 2.0 TOOLS

In this session will be represented the materials used for the development of this project.

### 2.1 Python

Python is an open source, object-oriented programming language. According to its own documentation, the language has high-level structures and efficient object orientation, together with a friendly and dynamic syntax, which results in good performance, and can be applied to various types of problems, such as web development, desktop development, business applications and scientific applications involving numbers and as it has libraries for data management and numbers like NumPy, it is widely used in machine learning (python.org).

### 2.2 Spyder

According to Spyder's website, the tool is "a powerful scientific environment written in Python, for Python, and designed by and for scientists, engineers and data analysts". The IDE provides a combination of advanced editing, analyzing, debugging, and profiling features of a comprehensive development tool with data exploration. The IDE also offers integration with some widely used packages such as NumPy, SciPy, Pandas, IPython, QtConsole, Matplotlib and SymPy, which are packets for number handling, and handling operations.

### 2.3 Keras

Keras is a high-level neural network library that runs over TensorFlow, which is an open source library for machine learning applicable to a wide variety of tasks. On its official website the API, which was written in Python, cites that it was developed with the aim of allowing experimentation and rapid prototyping supporting convolutional networks and allowing the use of processing via GPU.

Figure 1 - Most commonly used frameworks in 2018

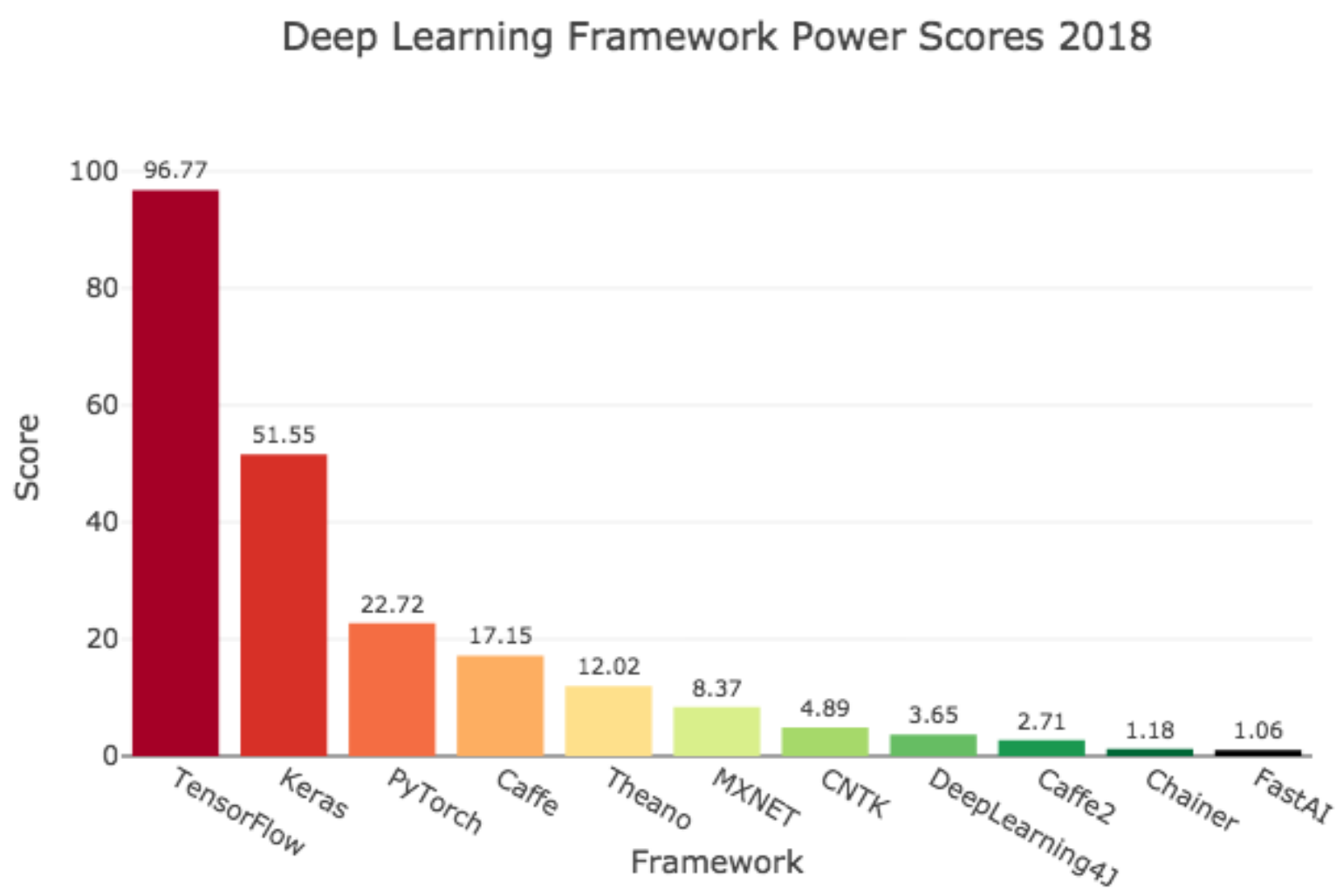

Source: keras.io

Figure 1 shows that Keras is one of the most commonly used tools for deep learning, and due to it, this framework was chosen to this work.

> With more than 250,000 individual users in mid-2018, Keras has a stronger adoption in the industry and research community than any other deep learning framework except TensorFlow itself.(KERAS. IO, between 2000 and 2019).

## 2.4 OpenCV

OpenCV is an open-source computer vision library. The library has more than 2,500 optimized algorithms, which include a comprehensive set of classic and advanced algorithms of computer vision and machine learning (OPENCV.ORG, 2019).

According to their documentation, the algorithms have the function of detecting and recognizing faces, identifying objects, classifying human actions in videos, tracking camera movements, tracking moving objects, extracting 3D models from objects, etc. According to SourceForge statistics by January 2019 OpenCV had more than 18 million downloads, being widely used in companies, research groups and government agencies.

The library has interfaces in C++, Python, Java and MATLAB and supports Windows, Linux, Android and Mac OS (OPENCV.ORG, 2019).

# 3.0 METHODS

In this topic will be represented the methods used for the development of this project.

## 3.1 Image Acquisition

For the work to be carried out, a training image base and a test image base are required to feed the application. The dataset was obtained by the institution EyePACS (Cuadros J.; Bresnik G.; EyePACS, 2009) which together with the California Health Care Foundation provide a public dataset with approximately 35,126 training images, covering from level 0 to level 4 of the disease, exactly for the purpose of research and development of solutions. However, only 708 belonged to the proliferative diabetic retinopathy class, which prevents the inclusion of the entire base to avoid class imbalance.

### 3.1.1 Class Imbalance

According to chart 1, the train dataset shows a large variation of examples among the 5 classes, being largely healthy examples in relation to the 708 images of the proliferative level of the disease, whose is the object of this work. This imbalanced proportion is exemplified by Murphey et al (2004, p. 117). The author exemplifies through a classification project among defective and non-defective examples of an automotive manufacturing environment. According to the author, it is common that in such classifications, we obtain a considerably larger number of healthy examples than defective, the result in practice is that the class with fewer images ends up being 'ignored', that is, the results tend to be belonging to the largest class due to the overwhelming difference in examples (MURPHEY et al (2004).

The dataset provided by EyePACS has the following number of images(Chart 1):

Chart 1 - Balance between training base classes

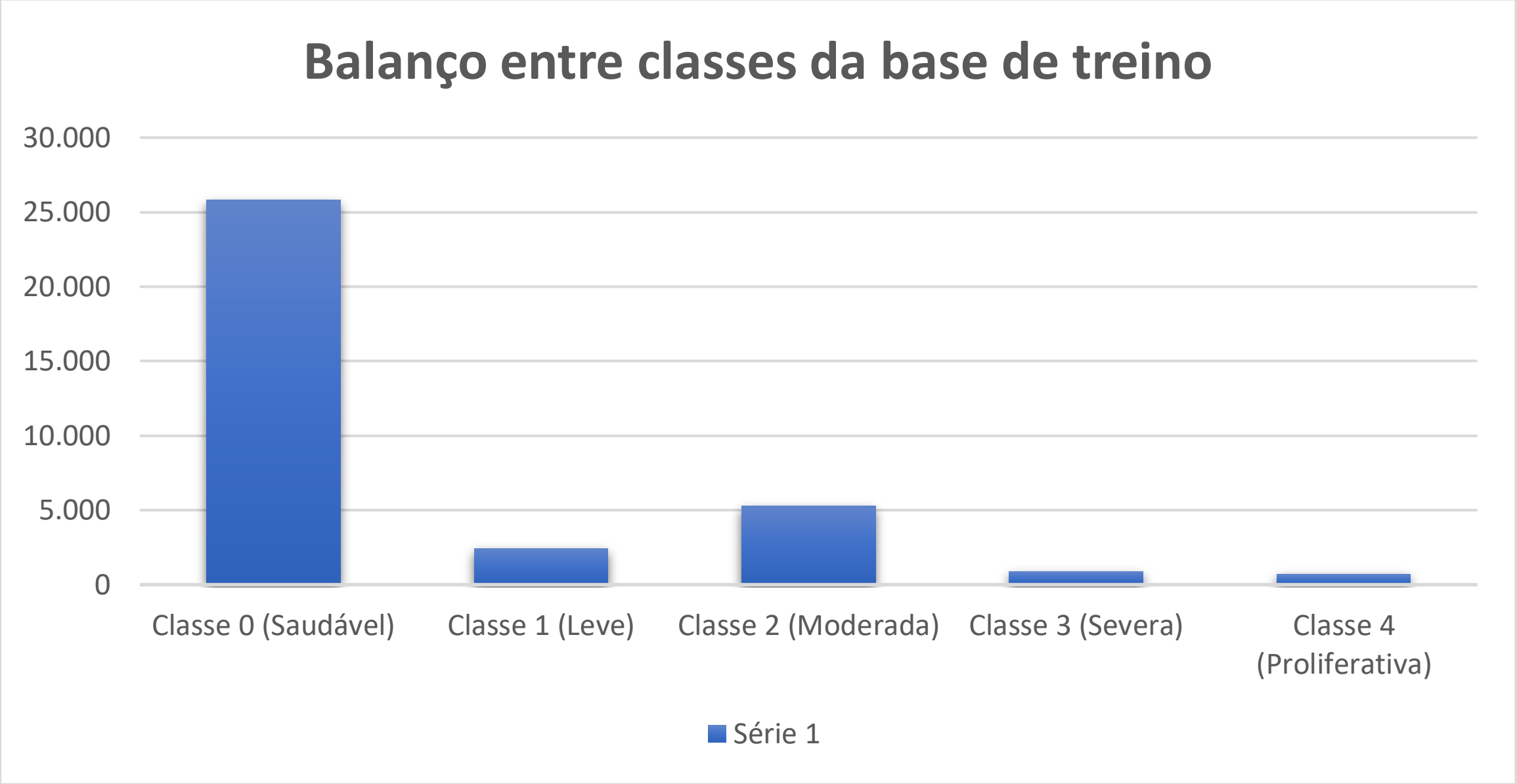

Source: Own author

To avoid class imbalance in this study, 700 PDR (Proliferative Diabetic Retinopathy) images and 700 healthy images were selected. The selection took place through the clarity of the photo, because there were some very low-quality images. Thus, 1400 images were obtained for training, 700 of each class (PDR and non-PDR).

For training validation, 72 images of each test base class were randomly selected, and for the final test, 50 images of each class present in the test base were selected.

### 3.1.2 Image Processing

The processing of the images that will be trained by some method seemed to be of great value to increase the accuracy of the results, either because a sharper image is obtained, or color manipulation. In this project, two forms of processing were included, which were the Gaussian blur effect and the CLAHE effect (Limited adaptive equalization of contrast histogram).

### 3.1.2.1 Gaussian Blur

According to the documentation of the OpenCV.org, the blur is obtained by convoluting the image through a given filter, being very useful for removing undesirable noises and edges.

Source Code 1 – Gaussian Blur

```
ImageCV[index] = cv2.addWeighted(
ImageCV[index],4,cv2.GaussianBlur(ImageCV[index],(0,0), 400/30),-4, 128)
```

Source: Own Author

The gaussian blur effect occurs by Source Code 1 and is useful for better reading of the image by the application, as it focuses on the retinal blood vessels. The problem is that in the dataset used for this work, there is a great variation of technological conditions in devices used to take the photo. This implies most images having camera artifacts and high illumination at the edges, details that the app may confuse with bleeding and leaks.

Figure 1 – Image without artifacts and blurs

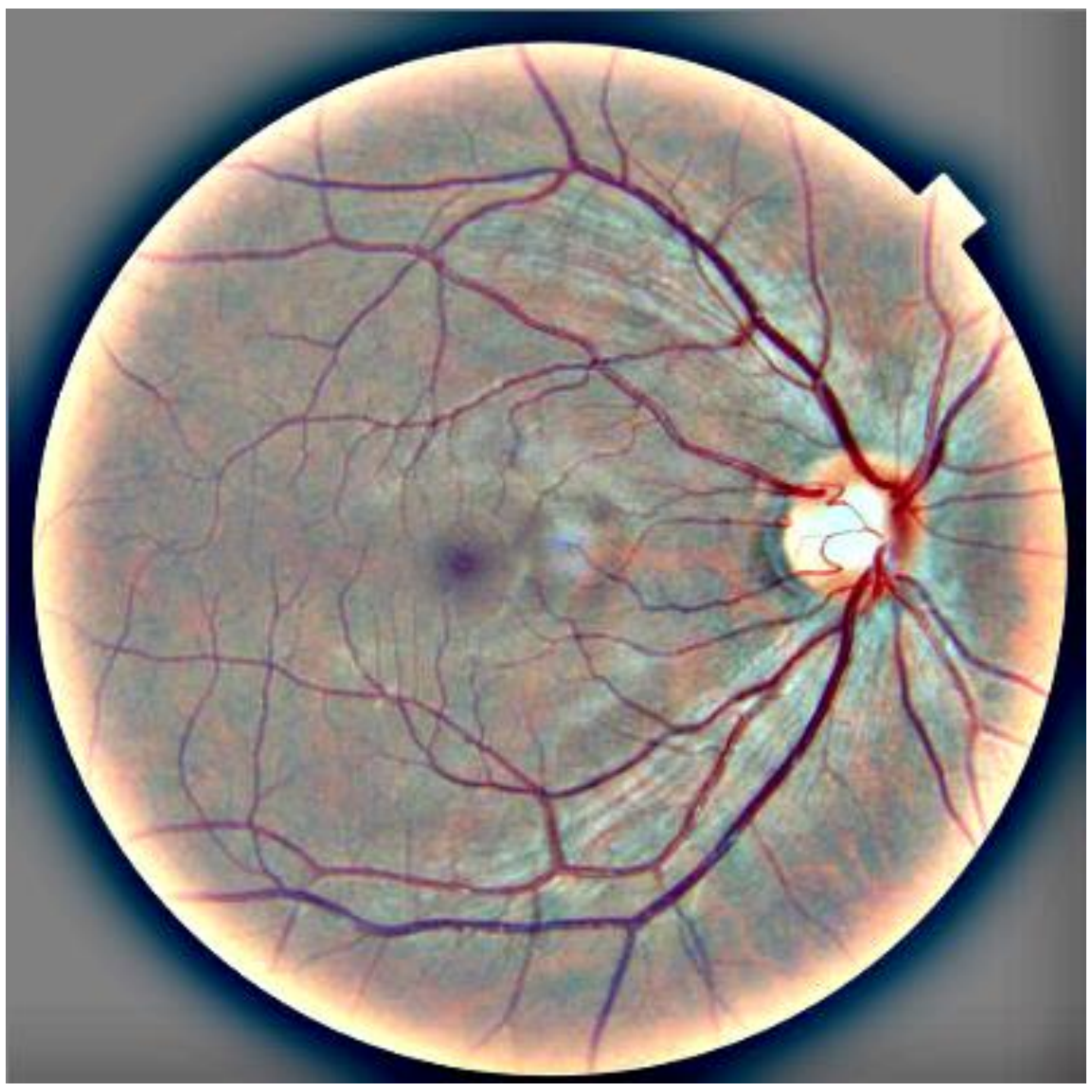

Source: topcon-medical.co.uk (between 2000 and 2019)

As can be seen in Figure 1 in an image with ideal conditions, the Gaussian effect is extremely advantageous to have a better analysis of blood vessels and general retinal condition.

Figure 2 - Image with camera artifacts (in red) present on the dataset

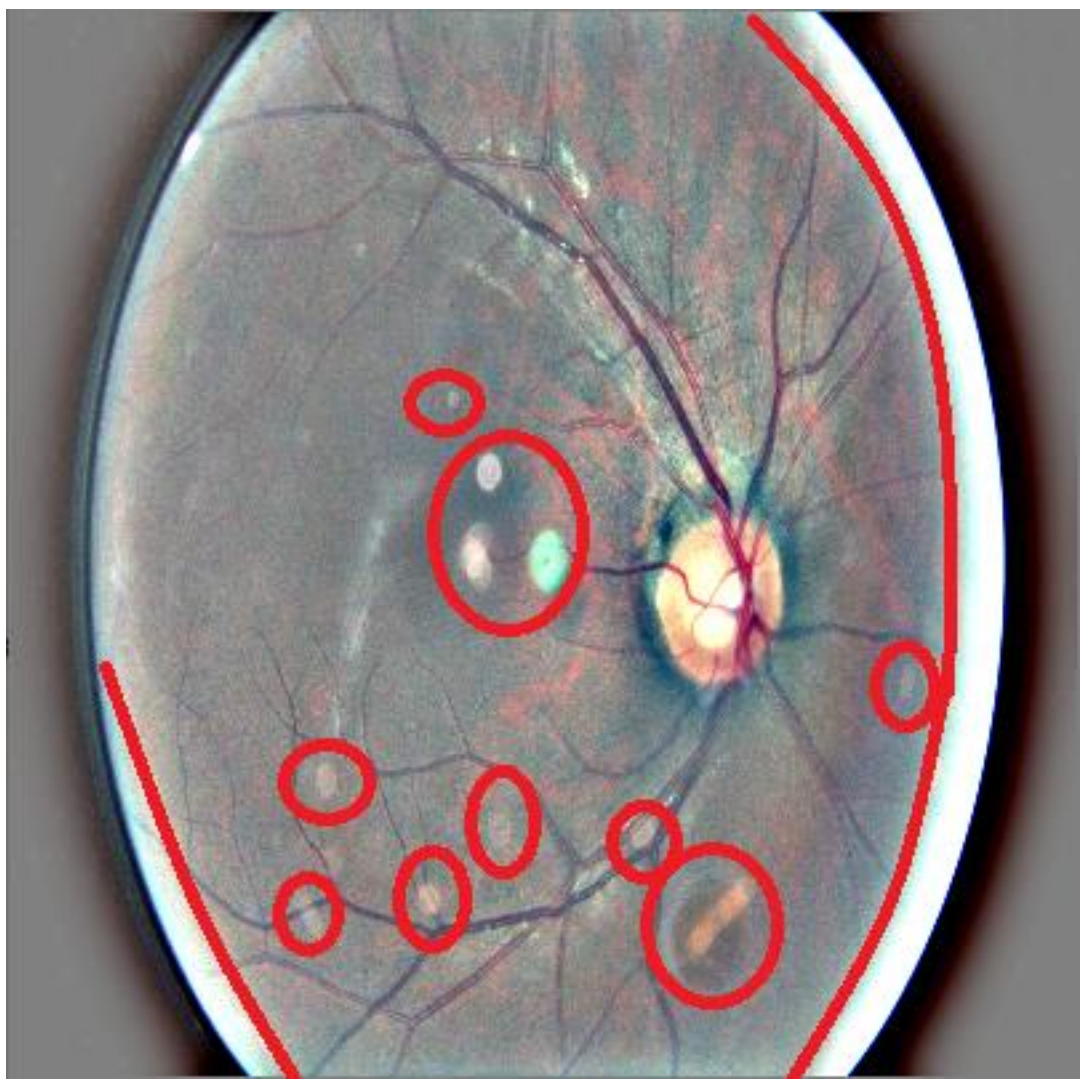

Source: Own Author (2019)

Figure 2 shows an image of the training dataset. A large number of artifacts and adverse lighting conditions are observed, which may be confused with some retinal alteration.

### 3.1.2.2 CLAHE Method

CLAHE is an adaptive equalization preprocessing method of histogram. According to the OpenCV documentation, this technique divides the image into frames called tiles that are equalized.

> ...Histogram would confine to a small region (unless there is noise). If noise is there, it will be amplified. To avoid this, contrast limiting is applied. If any histogram bin is above the specified contrast limit (by default 40 in OpenCV), those pixels are clipped and distributed uniformly to other bins before applying histogram equalization. After equalization, to remove artifacts in tile borders, bilinear interpolation is applied. (docs.opencv.org).

Figure 3 - Image processed with CLAHE

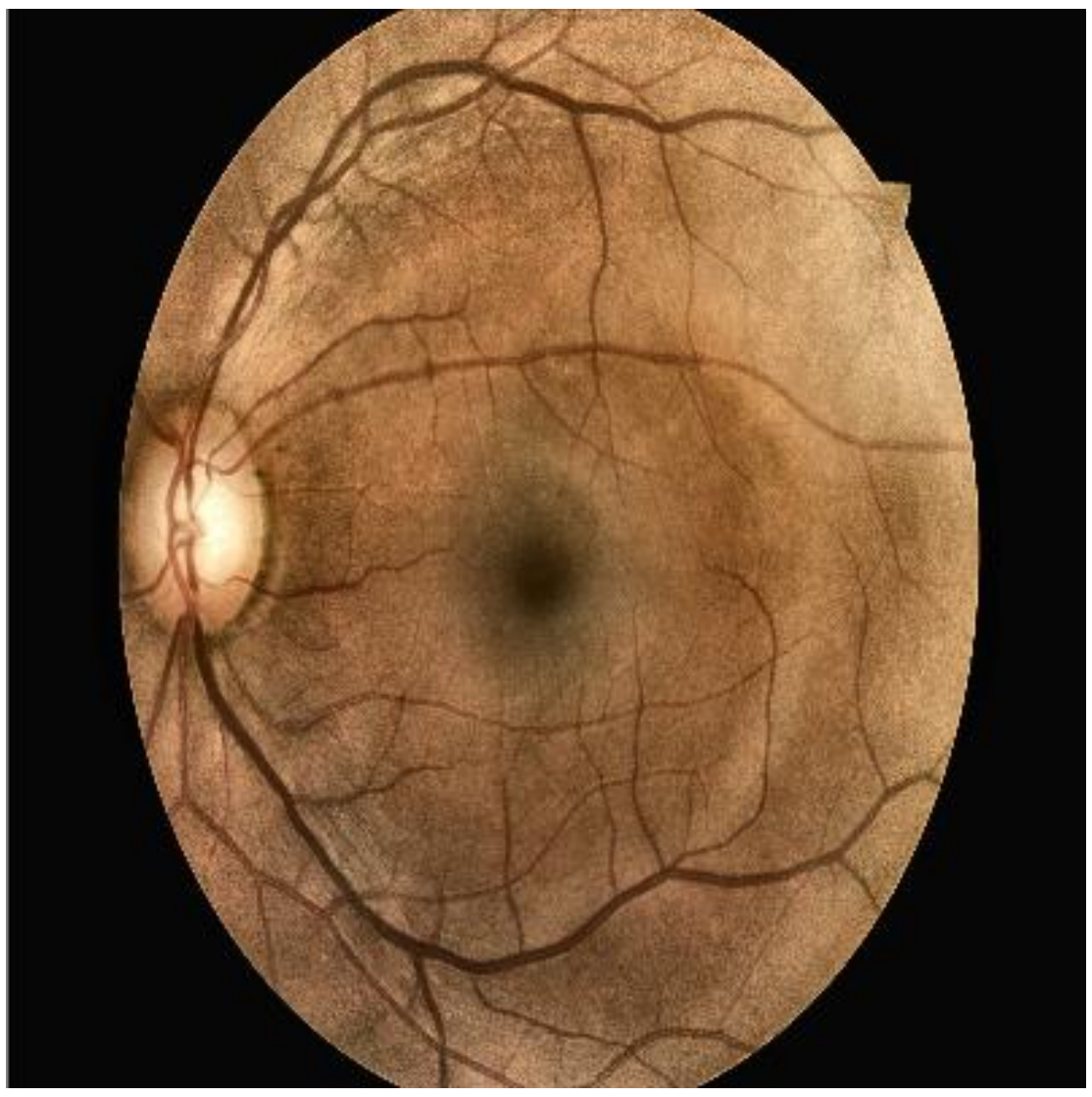

Source: Own Author (2019)

As seen in Figure 3, the CLAHE method really improves the contrast and visibility of retinal blood vessels, but just as the Gaussian effect, this method does not completely remove artifacts from the image.

In the final version of this project, therefore, no processing techniques were used.

### 3.1.3 Inclusion of images in the application

To include the training and validation images in the neural network, a file called train.py was created, which was responsible for the entire training. A method called ReadImages() has been created, which as its name suggests, reads the images that are in the training and validation directory, whose are passed as a parameter through the constants TRAIN_DIR and TEST_DIR, manipulates them and returns them in list form together with his class.

Both in the constant TRAIN_DIR and in the TEST_DIR there are two other subdirectories which contain the images to be used in the training process, which are nonPdr and pdr. These folders indicate the class belonging to the images that are inside.

Source Code 2 - ReadImages() Method

```
def ReadImages(Path):
    LabelList = list()
    ImageCV = list()
    classes = ["nonPdr", "pdr"]

    # Get all subdirectories
    FolderList = [f for f in os.listdir(Path) if not f.startswith('.')]
    # Loop over each directory
    for File in FolderList:
        for index, Image in enumerate(os.listdir(os.path.join(Path, File)):
      # Convert the path into a file
        ImageCV.append(cv2.resize(cv2.imread(os.path.join(Path, File)
                                                   + os.path.sep + Image), (224,224)))
        LabelList.append(classes.index(os.path.splitext(File)[0]))
    return ImageCV, LabelList
```

Source: Own Author (2019)

As set out in Source Code 2, the ReadImages() method creates two empty lists, which will later be fed with images and their classes. After storing the subdirectories in the FolderList variable, a loop was created that traverses all images by directory.

With the OpenCV library, the path of the image through which the loop is passing is read by the application using the cv2.imread method and then readjusted to the size of 224x224 px. The read and resized image is included in the ImageCV list, which will be returned by the method for all your calls, as well as LabelList, which the class of each image traversed by the loop is included.

For the method call 4 variables are created, one to store the images and one to store the labels for both testing and validation.

## 3.2 Neural Network Models

The architecture of a neural network consists of several layers with its neurons, such as input layers, convolution, dense, polling layers, etc...

> One feature of CNN networks applied to images is that the first layers are responsible for describing borders, intermediaries describe shapes, and the endings describe higher-level patterns, also known as semantic layers. Thus, when the database differs from that used in the training stage of the network (i.e., there is a need for *fine tuning),* the initial and intermediate layers usually bring better results for recognition, as they select more general characteristics (SANTOS et al, 2018).

Figure 4 – CNN Operation

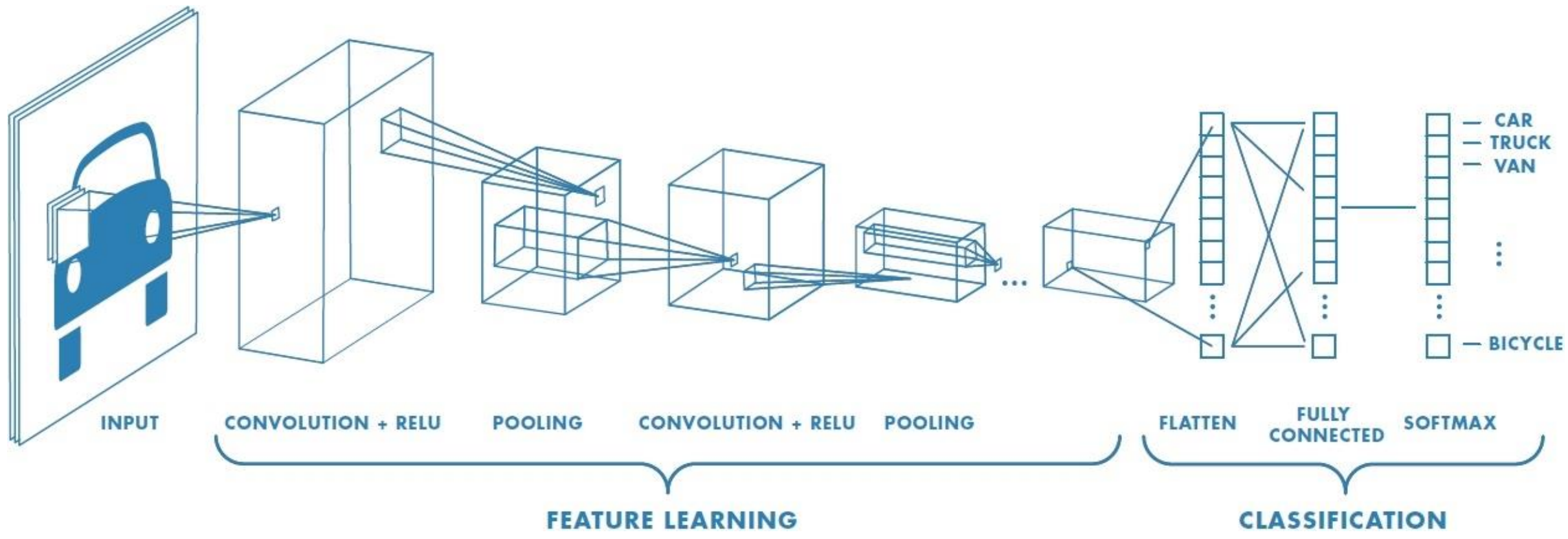

Source: MathWorks.com (between 2000 and 2019)

Figure 4 exemplifies the CNN(Convolutional Neural Network) operation, in which the image is 'disassembled' and each of its pixels pass through the layers of the network, being converted to RGB if the image is colored (figure 5), as is the case in this project, or in binary value if the image is grayscale (Stamford University).

Figure 5 - CNN Read Representation of a Color Image

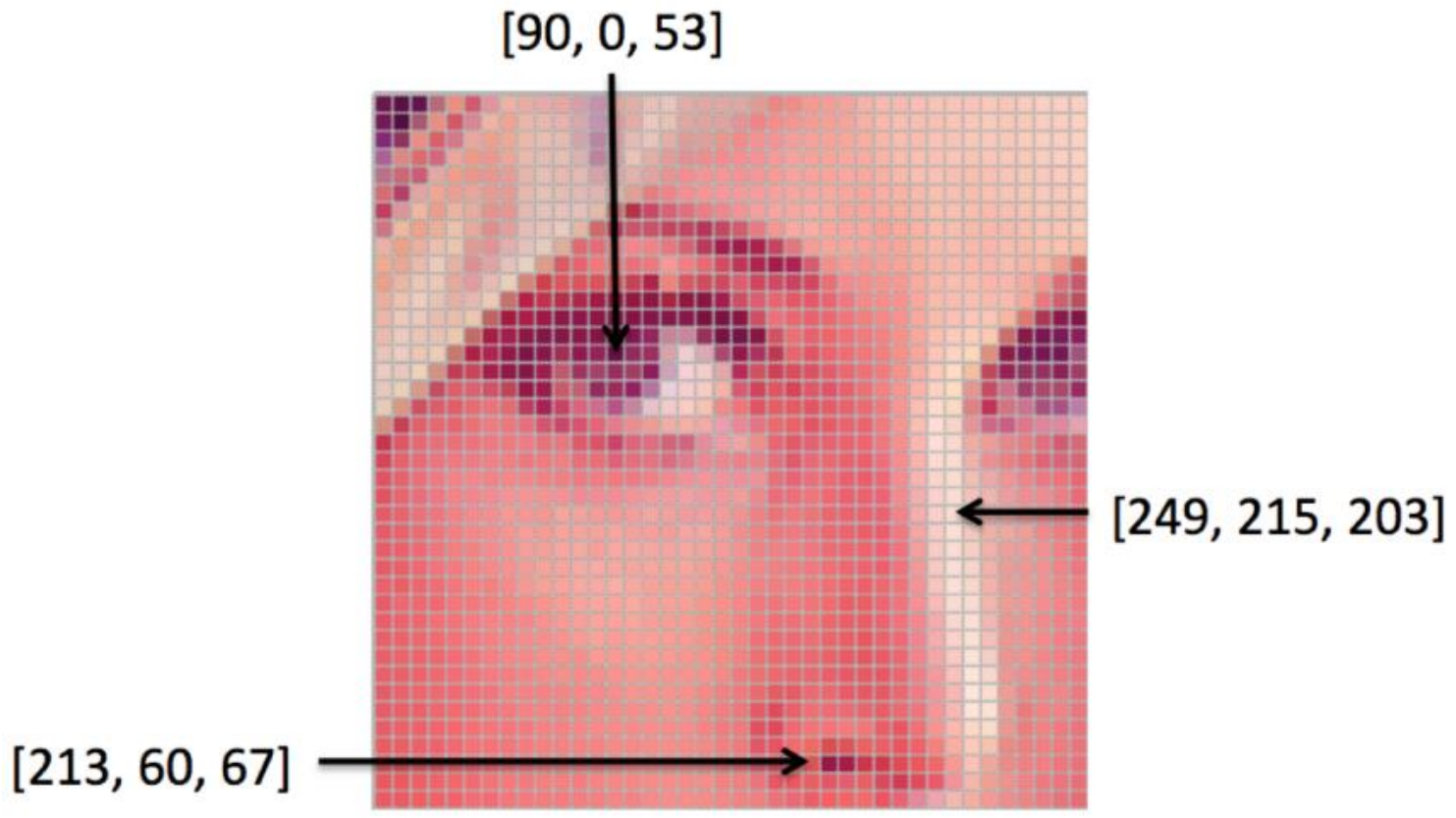

Source: Stamford University

According to Keras documentation, the user can create their own neural model through two ways that are: Sequential Model and the Functional API.

### 3.2.1 Sequential Model

The first attempt was to use the Sequential model. In this approach the user includes the layers of the model linearly.

Source Code 3 – Sequential Model

```
model = Sequential()
model.add(Conv2D(input_shape=(605,700,3), filters=64, kernel_size=(3,3), padding="same",activation="tanh"))
model.add(MaxPooling2D((2, 2)))
model.add(Flatten())
model.add(Dense(32, activation='tanh'))
model.add(Dense(1, activation='sigmoid'))
model.compile(optimizer='rmsprop', loss='binary_crossentropy', metrics=['accuracy'])

model.fit(np.array(data), np.array(labels), epochs=10, batch_size=16)

model.save('model.h5')
```

Source: Own Author (2019)

In source code 3 a standard convolutional model is specified using:

- A convolutional input layer specifying the size of the image, with 64 neurons and tahn activation function;
- A MaxPooling layer;
- A Flatten layer;
- Two dense layers, one output having 1 neuron and sigmoid activation, and the other with tahn activation with 32 neurons;

After adding the layers, the model.compile() method is called that according to keras documentation is responsible for "configuring the learning process", containing three arguments: An optimizer, a loss function, and a metric. In this example, rmsprop was used as an optimizer, binary_crossentropy as a loss function because the application has a binary scope (nonPdr or pdr), and for measurement the model accuracy was selected.

Once the model is created and compiled, it's time to train it with the *model.fit method(),* which receives the list of images, classes, the amount of times that the application will run, and the number of examples per epoch.

### 3.2.2 Functional *API*

The Functional API is a way to create more flexible models than Sequential: it can manipulate models with nonlinear topology, models with shared layers, and multi-input or output models (TensorFlow, 2018). In this model each layer is connected to the other, usually the previous one. The following template was created using Functional API:

Source Code 4 - Functional API Model

```
visible = Input(shape=(256,256,3))
conv1 = Conv2D(16, kernel_size=(3,3), activation='relu', strides=(1, 1))(visible)
conv2 = Conv2D(32, kernel_size=(3,3), activation='relu', strides=(1, 1))(conv1)
bat1 = BatchNormalization()(conv2)
conv3 = ZeroPadding2D(padding=(1, 1))(bat1)
pool1 = MaxPooling2D(pool_size=(2, 2))(conv3)
drop1 = Dropout(0.30)(pool1)
conv4 = Conv2D(32, kernel_size=(3,3), activation='relu', padding='valid', kernel_regularizer=regularizers.l2(0.01))(drop1)
conv5 = Conv2D(64, kernel_size=(3,3), activation='relu', padding='valid', kernel_regularizer=regularizers.l2(0.01))(conv4)
bat2 = BatchNormalization()(conv5)
pool2 = MaxPooling2D(pool_size=(1, 1))(bat2)
drop1 = Dropout(0.30)(pool2)
conv6 = Conv2D(128, kernel_size=(3,3), activation='relu', padding='valid', kernel_regularizer=regularizers.l2(0.01))(pool2)
conv7 = Conv2D(128, kernel_size=(2,2), activation='relu', strides=(1, 1), padding='valid') (conv6)
bat3 = BatchNormalization()(conv7)
pool3 = MaxPooling2D(pool_size=(1, 1))(bat3)
drop1 = Dropout(0.30)(pool3)
flat = Flatten()(pool3)
drop4 = Dropout(0.50)(flat)
output = Dense(1, activation='sigmoid')(drop4)
model = Model(inputs=visible, outputs=output)
opt = optimizers.adam(lr=0.001, decay=0.0)
model.compile(optimizer= opt, loss='binary_crossentropy', metrics=['accuracy'])
```

Source: Own Author (2019).

In source code 4, some changes were made regarding the sequential model, such as increased network, inclusion of the Dropout layer, decreased learning rate to 0.001, and the use of the Adam optimizer instead of rmsprop.

### 3.2.3 Overfitting

As stated by Cogswell et al (2016), overfitting is the situation when, in a neural model, it has a very high performance during training, however, a terrible accuracy in external data. Problem that is very susceptible when you have few examples of training, but can also occur on larger datasets depending on the situation.

According to Ying (2019), overfitting is "a fundamental issue in Machine learning, which prevents us from generalizing the models accordingly to adjust well the data observed in the training dataset, as well as in examples not seen by the application (test dataset)". Overfitting occurs due to the presence of noise, the limited size of the training dataset, and the complexity of the classifiers (YING, X. 2019). The author specifies some strategies to avoid overfitting in the neural network, including network reduction in order to reduce its complexity, the training dataset expansion, Early-Stopping, and regularization with Dropout layers. As claimed by Lemley et al (2017), the Data Augmentation is also a viable practice.

.

#### 3.2.3.1 Network Reduction

On the report of Ying (2019) eliminating some 'branches' of the network is a significant theory used to reduce the complexity of classification by removing less significant or irrelevant data and avoiding overfitting. Especially on smaller datasets, a leaner network helps contain the loss.

This practice was carried out in this project, eliminating dense layers and reducing the network shape.

#### 3.2.3.2 Dataset Expansion

In machine learning, the algorithm is not the only key that affects the final classification accuracy. Its performance can be significantly affected by the quantity and quality of the training dataset, especially when it comes to supervised learning (YING, 2019).

For the author, model training is a process of adjusting its parameters, because well-adjusted parameters strike a good balance between the accuracy and regularity of the training and then inhibit overfitting, and to adjust these parameters, the model needs enough samples for learning (YING, 2019).

In this project, the database used has 1400 images for training, and following the author's logic, if it could get more examples, the greater accuracy and less overfitting.

#### 3.2.3.3 Early Stopping

Raskutti et al (2014) explains early stopping as being "... a form of regularization based on the choice of when to stop running an iterative algorithm." The author explains that at each step of an iterative algorithm will reduce bias, but will increase variation, so that early stopping ensures that the variation is not too high, that is, when greater error variation in the processing of the time is detected, the algorithm stops executing.

In *Keras* this practice can be performed as follows (source code 5):

Source Code 5 - Early Stopping call

```
from keras.callbacks import EarlyStopping

es = EarlyStopping(monitor='val_loss', verbose=1)
```

Source: Own Author (2019)

The loss function was used to be the method monitor (source code 5), so every time the loss value increases (signaling overfitting), the application terminates its workout. The variable 'es' is later passed with argument of the method fit_generator.

#### 3.2.3.4 Dropout

According to Ying (2019) Dropout is a popular and effective technique against overfitting in neural networks. The initial idea of Dropout is to randomly discard units and connections from neural networks during training. This prevents units from adapting too much.

Figure 6 – Normal Neural Network (left) / After *Dropout* implementation (right)

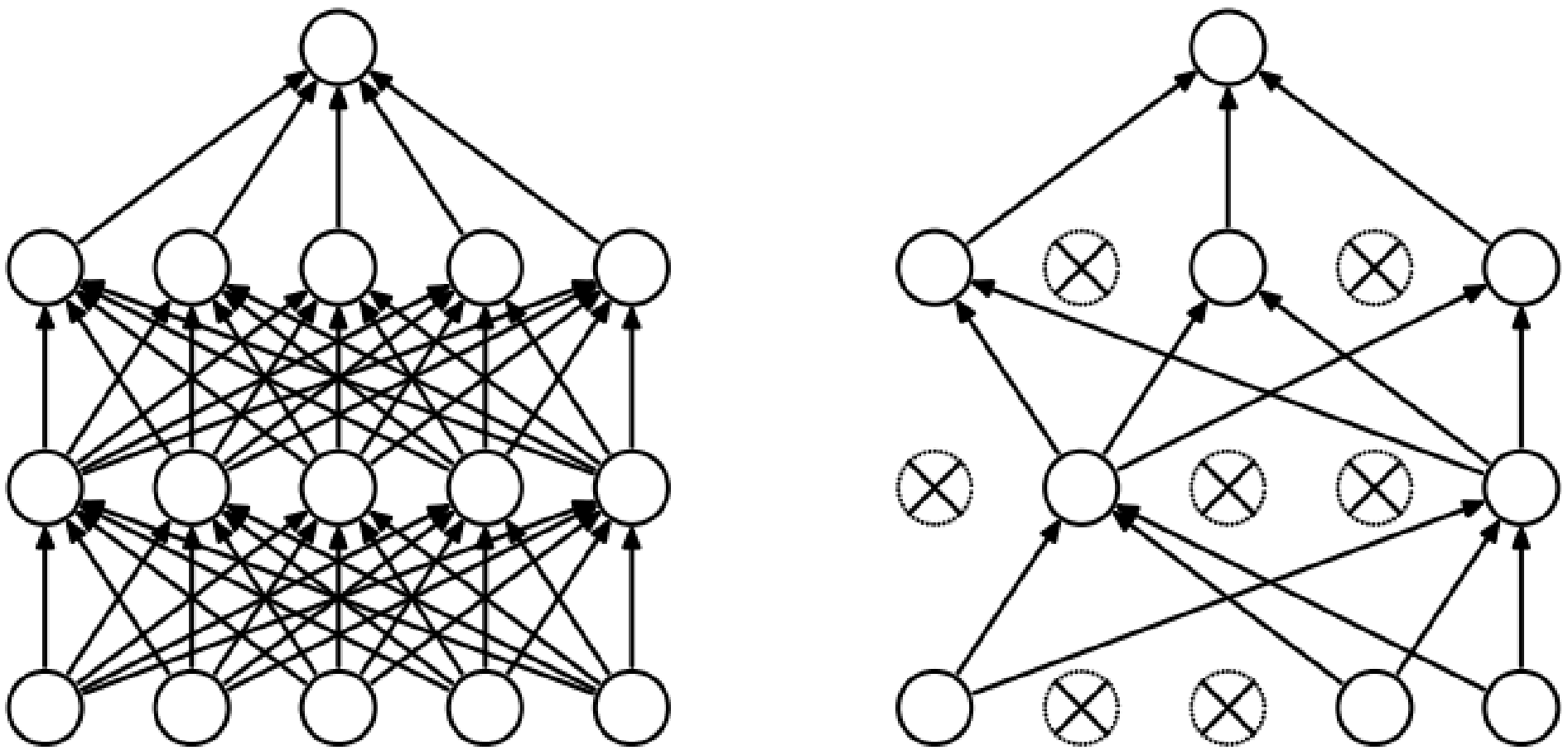

Source: SRIVASTAVA et al (2014)

Figure 6 shows the Dropout effect on the neural network, 'cleaning' units randomly, which reduces the network (SRIVASTAVA et al, 2014). The process that this implementation aims to disassemble is the co-adaptation that occurs "when two or more hidden units depend on each other to perform some function that helps adjust the training data, becoming highly correlated." (SRIVASTAVA et al, 2014).

The Dropout implementation in Keras took place through the following code:

Source Code 6 – Dropout Implementation

```
from keras.layers.core import Dropout

base_out = Dropout(0.60)(base_out)
```

Source: Own Author (2019)

As source code 6 shows, a 0.6 Dropout rate was applied, this rate represents the fraction of the units that will be disregarded (keras.io, 2019).

#### 3.2.3.5 Data Augmentation

The practice of data augmentation serves as a kind of regularization, reducing the chance of overfitting extracting more general information from the database and passing it to the network (LEMLEY et al, 2017). Generally, the practice is performed by trial and error, and the types of augmentations are limited to the researcher's imagination, time and experience. (LEMLEY et al, 2017).

Keras has a method for making data augmentation in training called ImageDataGenerator.

Source Code 7 - Data Augmentation implementation

```
from keras.preprocessing.image import ImageDataGenerator

datagen = ImageDataGenerator(
    rotation_range=30,
    zoom_range=0.15,
    featurewise_std_normalization=True,
    width_shift_range=0.2,
    height_shift_range=0.2,
    shear_range=0.15,
    horizontal_flip=True,
    fill_mode="nearest")

datagen.fit(data)
mean = datagen.mean
std = datagen.std

print(mean, "mean")
print(std, "std")
```

Source: Own Author (2019)

Source code 7 shows the augmentation made in training, using techniques such as image rotation, zoom, and normalization. After the process is carried out, it is necessary to normalize the examples, which is done by taking the datagen mean value and variance (embedded with the augmentation techniques). This values are restored in the file test.py, and this normalization is applied to the test samples, which was done with the following code:

Source Code 8 - Normalization

```
def normalize(x, mean, std):
    x[..., 0] -= mean[0]
    x[..., 1] -= mean[1]
    x[..., 2] -= mean[2]
    x[..., 0] /= std[0]
    x[..., 1] /= std[1]
    x[..., 2] /= std[2]
    return x

#Values generated during train
x = normalize(x, [0.12803958, 0.1789845,  0.23866335],
                [0.14304826, 0.18132521, 0.23619336])
```

Source: Own Author (2019)

In source code 8, the normalization performed in the test.py file is specified, responsible for testing the model on external images. This practice was needed because the original model normalization would be based on the images in which it was originally trained, in this case, ImageNet. However, in this case it was necessary to include the mean and variance values of the train dataset, therefore, after displaying the values on the screen, in the training period, they were passed as parameters for the function called normalize that focused on the images of test dataset.

### 3.3 Transfer Learning

As maintained by Fawaz et al (2018) the practice of transferring learning, is the process of first training a neural network based on a source, and then transferring the characteristics learned by the network (called weights), to a second network that will be trained on another dataset.

Figure 7 – Transfer learning process example

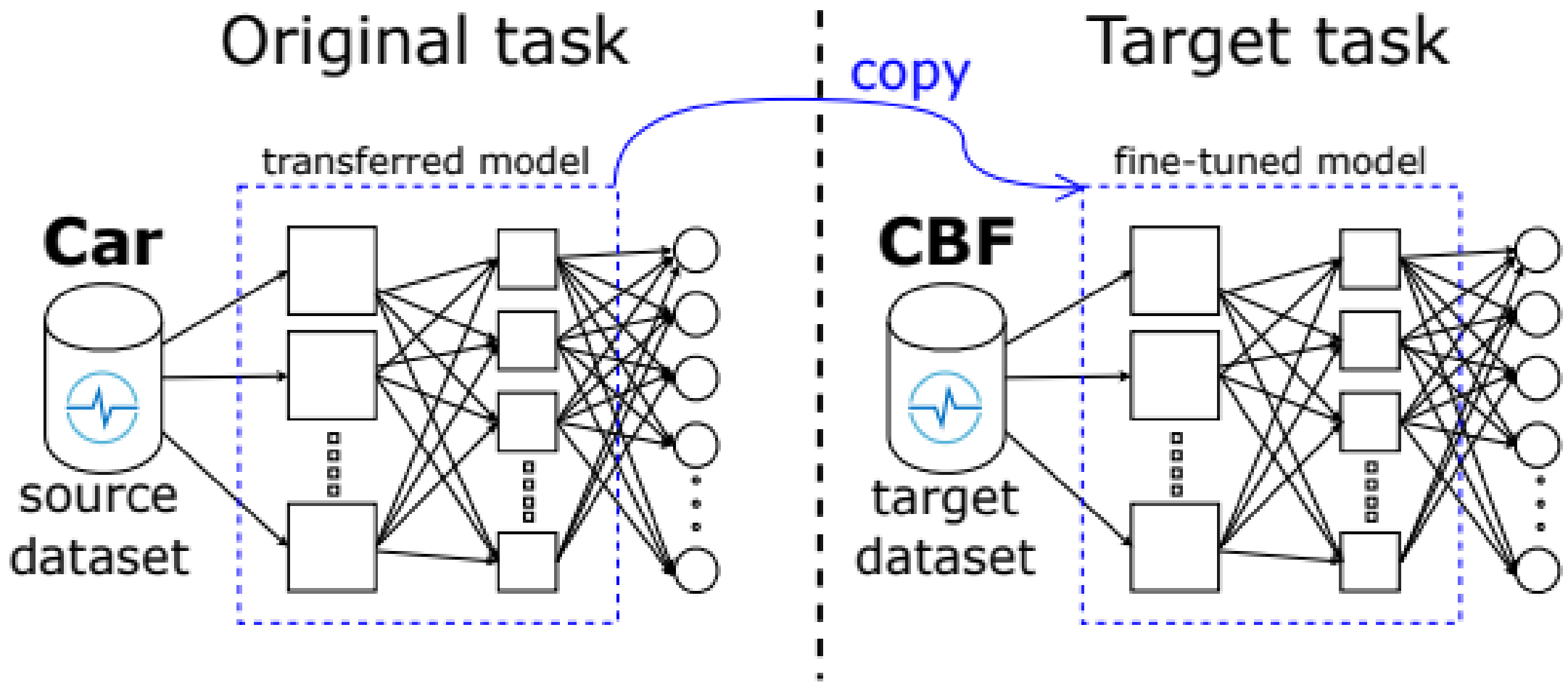

Source: FAWAZ et al (2018)

Figure 7 shows an example of transfer learning, which in the first model called 'Car' (source dataset) is trained and its weights are applied through fine-tuning in the target dataset.

This practice fits into this project because, according to Yosinski et al (2014) "When target dataset is significantly smaller than the base dataset, transfer learning can be a powerful tool to allow training of a large neural network without presenting overfitting".

The usual approach to transfer learning is to train a base network and then copy its first n layers in the first n layers of a target network. The remaining layers of the target network are randomly initialized and trained for the target task. (YOSINSKI et al, 2014).

ImageNet is a database with more than 14 million images used to assist researchers with an accessible database (Image-Net.org ,2010). Keras fully supports some pre-trained models such as: Xception, VGG16, VGG19, ResNet, ResNetV2, InceptionV3, InceptionResNetV2, MobileNet, MobileNetV2, DenseNet and NASNet (keras.io, 2019).

Due to the effort to avoid overfitting, the smaller, or simpler a model, the less likely to occur overfit. According to Lundström (2017), the VGG-16 is a very popular model precisely because it has a simple architecture.

> VGG models are considered simple due to two reasons. The first is the use of 3x3 convolutions across the network. The second is the duplication of the number of resource maps after max-pooling of 2x2. 2x2 max-pooling is also used across the network. These simple choices eliminated the need to adjust convolutional filter sizes and individual layer sizes. LUNDSTRÖM, D. (2017).

The VGG16/19 model is among the top 5 models, which achieve the highest accuracy of image classification (BAJIĆ et al, 2019). There is a competition provided by ImageNet that evaluates algorithms for large-scale object detection and classification. Competition in which, in the 2014 edition, vgg-16, among several other models, came first in the localization competition, and second in the classification competition (image-net.org, 2014).

### 3.3.1 VGG-16

According to Keras documentation, the following code is sufficient to include the VGG-16 structure in a local model:

Source Code 9 - VGG-16 Initialization

```
from keras.applications.vgg16 import VGG16

vgg16_model = VGG16(weights="imagenet", include_top=False)
```

Source: Own Author (2019)

Source code 9 assigns the 'vgg16_model' variable to the pre-trained VGG-16 model with ImageNet weights.

Figure 8 - VGG-16 Original Structure

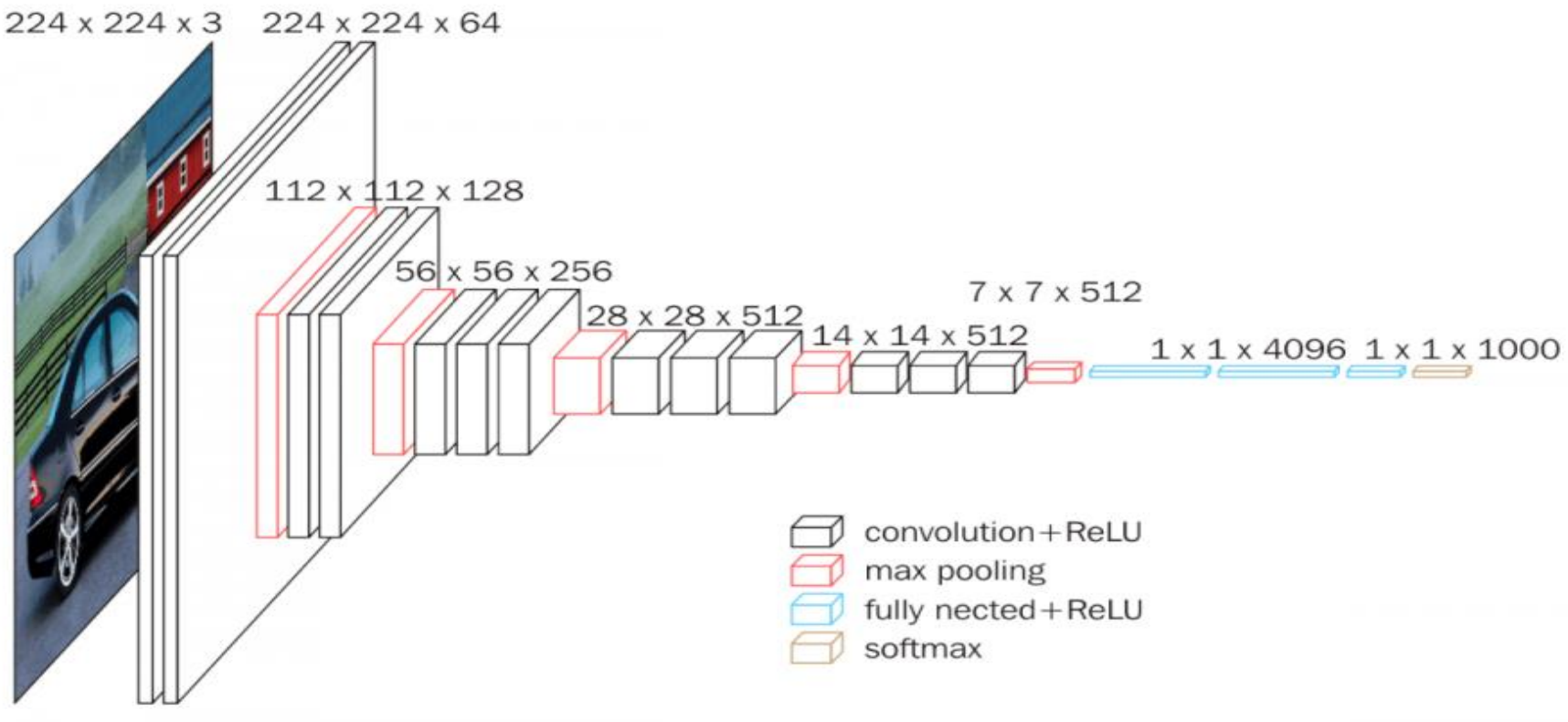

Source: HASSAN, M. (2018)

Figure 8 specifies the original structure of VGG-16, which, as can be seen, has 13 convolutional layers, which as stated by Reyes et al (2015), "... can be understood as filter banks that transform an input image into another image, highlighting specific patterns.", followed by 5 pooling layers, described by Romanuke (2017) as responsible for progressively reducing the spatial size of the representation. Thus, it reduces the number of parameters and the amount of computing on the network also assisting in to avoid overfitting. In the final part, there are 3 dense layers – or fully connected layers – whose take a vector as input and produce another vector as output (REYES et al, 2015). and as an output layer, a softmax with 1000 neurons representing its 1000 classes.

To perform the transfer learning, the following code snippet was used.

Source Code 10 – Fine-Tuning process

```
vgg16_model = VGG16(weights="imagenet", include_top=False)

base_model = Model(input=vgg16_model.input,
                   output=vgg16_model.get_layer("block5_pool").output)

#attach a new top layer
base_out = base_model.output
base_out = Dropout(0.60)(base_out)
base_out = Reshape((25088,))(base_out)
fc2 = Dropout(0.60)(base_out)

top_preds = Dense(1, activation="sigmoid")(fc2)

for layer in base_model.layers[0:14]:
    layer.trainable = False

model = Model(input=base_model.input, output=top_preds)

adam = Adam(lr=1e-4)
model.compile(optimizer=adam, loss="binary_crossentropy", metrics=["accuracy"])

model.fit_generator(datagen.flow(data, np.array(labels), batch_size=32),
                            steps_per_epoch=len(data) / 32, epochs=10,
                           validation_data=(valid, np.array(vlabels)),
                            nb_val_samples=72, callbacks=[es])
```

Source: Own Author (2019)

In source code 10 the fine-tune process with VGG-16 is specified. As the parameter *include_top = false* was passed, the last 3 dense layers of the original model (VGG-16) do not remain in the new model, so in the variable *base_model*, the model was stored having as input, the normal layer of input, and as output, the last layer of the model, which in this case is the layer of MaxPooling called *block5_pool*.

After storing the model base, 4 more layers were included in the network. Two Dropout layers with 0.60 rate with a Reshape layer in the middle, which in this case is flattening the pixels at 25088 parameters linearly.

The last layer, classification, was included with 1 neuron having sigmoid activation. Thus, the result of class predictions will be a value between 0 and 1, being 0 the first class, in this case 'nonPdr' and 1 the second class 'pdr'.

Next, the weights of the first 14 layers are frozen, so that only the last 3 convolutional layers are trained. Soon after the model is updated with the new output layer and builds, the learning rate was reduced, because according to Wilson et al (2001) "Decreasing the learning rate can significantly improve the accuracy of generalization, especially in large and complex problems." Therefore, the 0.0001 learning rate was used in the project.

Finally, the *method fit_generator* is called. In it passes the variables that store the examples and classes of training and validation, as well as the Callback of Early Stopping that was developed previously. The number of 10 epochs is set to run the training. Also including transformations performed by *data augmentation* through the *datagen.flow* method.

Table 1 - Layers used to fine-tuning

| LAYER | SHAPE |
|---|---|
| Input | 224x224x3 (WxHxRGB) |
| Conv 1-1 | 224x224x64 |
| Conv 1-2 | 224x224x64 |
| Max Pooling | 112x112x64 |
| Conv 2-1 | 112x112x128 |
| Conv 2-2 | 112x112x128 |
| Max Pooling | 56x56x128 |
| Conv 3-1 | 56x56x256 |
| Conv 3-2 | 56x56x256 |
| Max Pooling | 28x28x256 |
| Conv 4-1 | 28x28x512 |
| Conv 4-2 | 28x28x512 |
| Conv 4-3 | 28x28x512 |
| Max Pooling | 14x14x512 |
| Conv 5-1 | 14x14x512 |
| Conv 5-2 | 14x14x512 |
| Conv 5-2 | 14x14x512 |
| Dropout | 0.60 |
| Reshape(Flatten) | 25088 |
| Dropout | 0.60 |
| Output - Dense | 1 |

Source: Own Author (2019)

Table 1 is the representation of the network created with *fine-tuning* of the VGG-16. It is noted the removal of the dense layers and classification of the original model, replacing with a binary classification, as santos et al (2018) quotes "... only the convolutional part of the network and not its classifier can be reused, since you can choose a fully connected network with a different number of neurons." The output layer that originally had softmax activation, has now been changed to sigmoid with only one neuron, which will result in a prediction with a value between 0 and 1, representing the identification of the first class (nonPdr) for predictions up to 0.5, or the second class (pdr) for predictions from 0.5 to 1.0.

The choice of two layers of Dropout was due to the incidence of overfitting, practice that, according to Hinton et al (2012) "can be reduced using Dropout avoiding complex co-adaptations in the test base".

Figure 9 – Model training

```
[[[0.12803958 0.1789845  0.23866335]]] mean
[[[0.14304826 0.18132521 0.23619336]]] std
C:/Users/Junior/Desktop/DR/train.py:105: UserWarning: The semantics of the Keras 2 argument
`steps_per_epoch` is not the same as the Keras 1 argument `samples_per_epoch`.
`steps_per_epoch` is the number of batches to draw from the generator at each epoch. Basically
steps_per_epoch = samples_per_epoch/batch_size. Similarly `nb_val_samples`->`validation_steps`
and `val_samples`->`steps` arguments have changed. Update your method calls accordingly.
  nb_val_samples=72, callbacks=[es])
C:/Users/Junior/Desktop/DR/train.py:105: UserWarning: Update your `fit_generator` call to the
Keras 2 API: `fit_generator(<keras_pre..., steps_per_epoch=43.78125, epochs=10,
validation_data=(array([[[..., callbacks=[<keras.ca..., validation_steps=72)`
  nb_val_samples=72, callbacks=[es])
Epoch 1/10
44/43 [==============================] - 472s 11s/step - loss: 0.7611 - acc: 0.6495 -
val_loss: 0.6714 - val_acc: 0.5694
Epoch 2/10
44/43 [==============================] - 474s 11s/step - loss: 0.4944 - acc: 0.7767 -
val_loss: 0.6663 - val_acc: 0.6111
Epoch 3/10
 3/43 [=>............................] - ETA: 7:02 - loss: 0.6395 - acc: 0.7292
```

Source: Own Author (2019)

Figure 9 shows the model in the training phase, with the print of the mean and variance for later inclusion in the test.py. The training process usually goes on without overfitting is also observed, with the loss decreasing and the accuracy increasing between epochs.

# 4.0 RESULTS AND DISCUSSION

In the process of predicting classes we have the incidence of the previously trained model on the test base. The model is loaded using the native method *load_model,* which recreates and compiles the model for prediction.

Test images are obtained in the same way as training images and after passing through normalization, the *model.predict()* method is performed, which returns an input-based prediction between classes programmed between 0 (nonPdr) and 1 (pdr), it is then assumed that any prediction above 0.5 is considered class 1.

At first the default layer creation structure was used and since then the model has overfitting, returning the same value for all test images. After attempting to use the Functional API the results remained the same in whatever the image, even after the inclusion of Dropout and Data Augmentation layers, so the transfer learning process with VGG-16 were used.

Using the Transfer learning model with both *Gaussian Blur* and CLAHE, they result in overfit in season two. Apart from any type of processing in the images the model trained appropriately without presenting overfit.

The application test took place through 100 test images. Of these 100, 50 of each class, i.e. 50 retinas with Proliferative Diabetic Retinopathy (PDR) and 50 with other levels, or without the disease. All these images had their diagnosis ensured by the Ophthalmologist Eduardo Menezes.

Figure 10 - Retina without Proliferative Diabetic Retinopathy

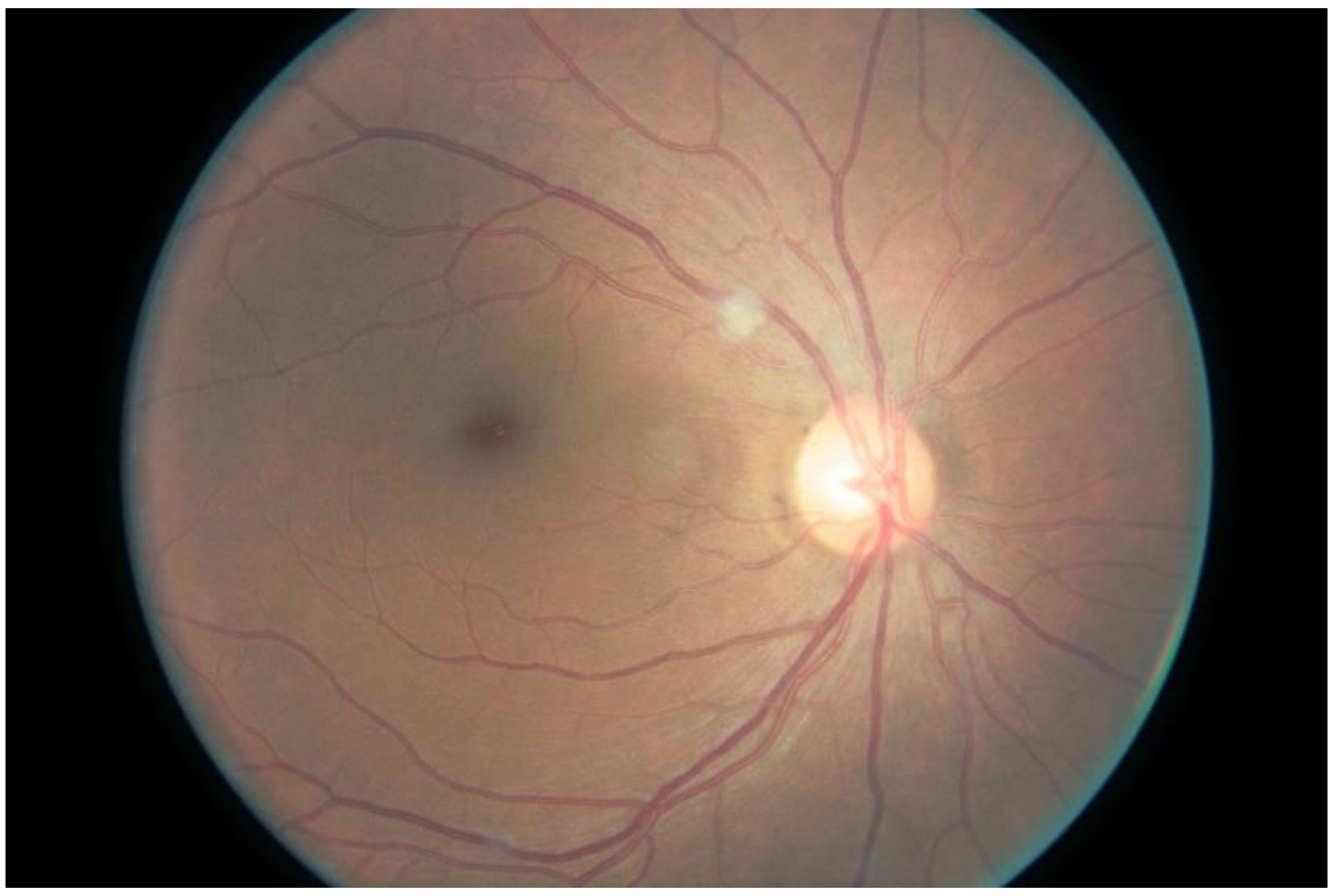

Source: CUADROS J.; BRESNIK G.; EyePACS (2009)

Figure 10 shows an example of retina without PDR, which, after the application of the neural model returned a value of 0.10658325, which is considered a low value and consistent with the level of the disease presented.

Figure 11 - Retina with Proliferative Diabetic Retinopathy

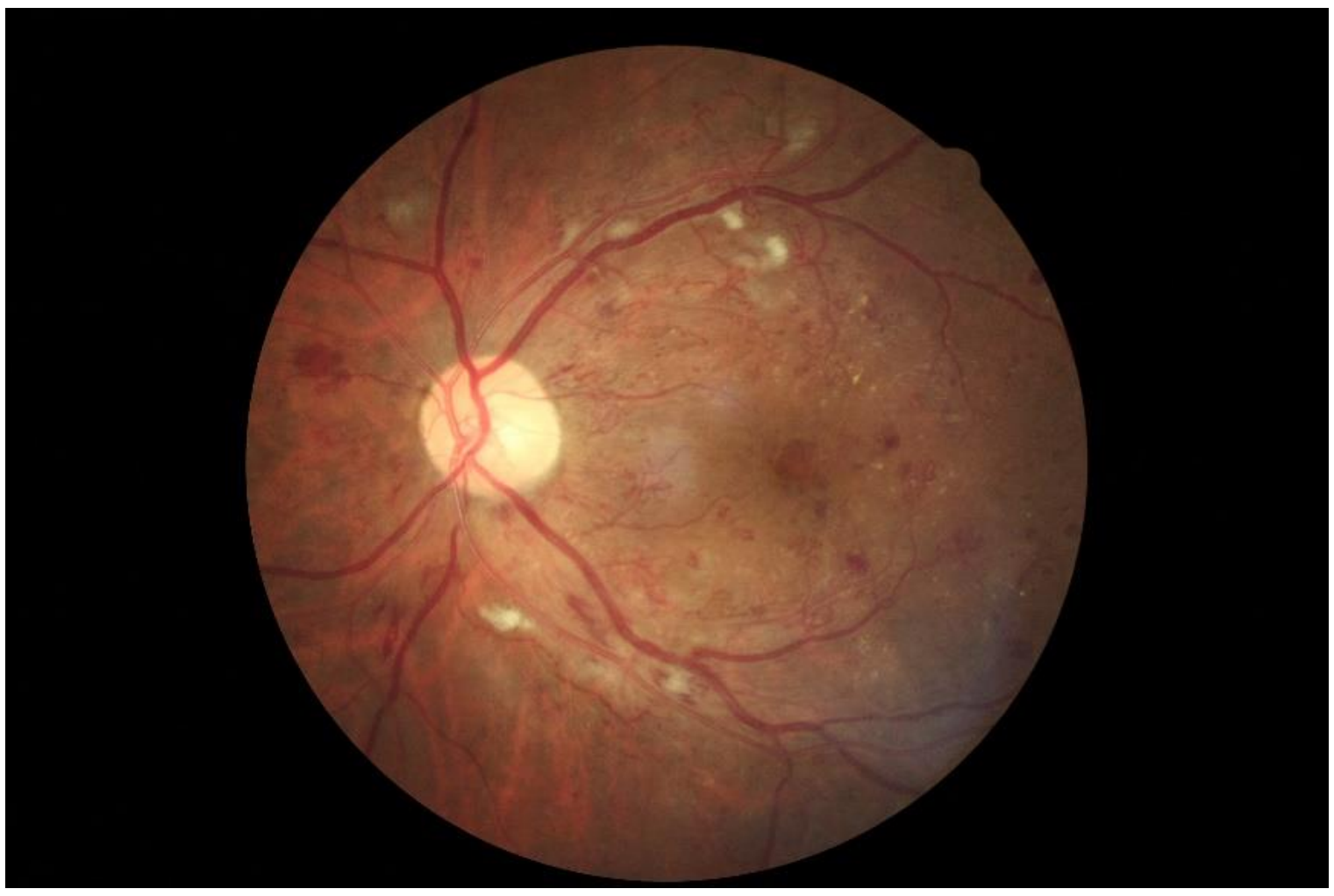

Source: CUADROS J.; BRESNIK G.; EyePACS (2009)

Figure 11 presents a retina example with proliferative diabetic retinopathy, which, after the application of the neural model returned a value of 0.9691823, very close to 1.0, number representing class 1 ('pdr').

Table 2 - Confusion Matrix

| | **Predicted "pdr"** | **Predicted "nonPdr"** |
|---|---|---|
| **"pdr" Actual** | 41 (TP) | 9 (FP) |
| **"nonPdr" Actual** | 9 (FN) | 41 (TN) |

Source: Own Author (2019)

Table 2 displays the confusion matrix obtained after evaluating the application test results. It is noticed that there was an equal proportion of results. Among the 50 images with PDR, 41 true positives (PV) and 9 false negatives (FN) were obtained, that is, 41 hits and 9 errors. Similarly, the 50 images without PDR (nonPdr), the result was 9 false positives (FP) and 41 true negatives (VN), resulting in 41 hits and 9 errors as well.

To calculate the accuracy of the classification follows the following formula (ASHCROFT, Mike. 2016):

$$\frac{TP + TN}{TP + TN + NP + NN}$$

In view of this, we found an accuracy of 82% in the model. As the results of the classification were symmetrical, the possibility of calculating other metrics such as recall, f-measure, and accuracy was dispensed, considering that all calculations would return to the same percentage of 82%.

According to the ophthalmologist Eduardo Menezes, the accuracy of 82% is a good initial indicator, which, however, should be improved to ensure the assertiveness of the diagnosis.

Figura 12 – Some predictions on the test dataset with PDR retinas

```
PDR >>> 853_left.jpeg
[[0.670009]]
PDR >>> 8639_left.jpeg
[[0.9227914]]
PDR >>> 8639_right.jpeg
[[0.9967681]]
PDR >>> 9559_right.jpeg
[[0.9427955]]
Number of retinas with PDR:  41
Number of retinas without PDR:  9
```

Fonte: Próprio autor (2019)

Figure 24 display some predictions obtained through the test dataset with PDR. The value between brackets represents the range of 0 and 1, representing both classes, being any prediction above 0.5, considered class 1, or PDR.

## 5.0 CONCLUSION

According to what was observed in the development and research process of this project, it is concluded that the aid of imaging technology is little explored in ophthalmologic offices. A process that, if applied, can generate greater assertiveness and agility to the diagnostic process of a disease as serious as is diabetic retinopathy, the sooner as possible to start treatment, the better the chances of preserving or softening the damage to the vision of the Patient. With this application we were able to reach the population in the places farther from the large centers, greatly reducing the sequelae that diabetic retinopathy can cause (Eduardo Henrique Marques Menezes).

The practice of applying this diagnostic method was received with great acceptance by ophthalmologist Eduardo Henrique Marques Menezes, who followed the development process. According to him "The development of a software to identify signs of Diabetic Retinopathy is extremely important for the identification and treatment of this disease that is one of the main causes of blindness in Brazil and in the world".

Considering that for a machine learning application a large number of examples are recommended, especially in this case where anomalies are thorough, if more examples are achieved, the result could be much better. Adverse conditions of luminosity and artifacts present in the images may have negatively influenced during training. Overfitting, on the other hand, prevented the inclusion of techniques for image processing.

The use of technologies for process automation is a positive practice that can have great benefits for both the doctor and the patient facilitating the early and agile detection of the disease, including acting as a support to the medical decision.

Like every intelligent application directed to health, there is a great learning in the medical field, understanding more about the disease, in this case, and acting for the efficiency and well-being of the community, eliminating time-consuming, unsafe processes, and in some inaccessible cases.

## 6.0 FUTURE WORK

Task automation is a growing practice in the world. It is of great importance the decrease in time, money and resources that you have to cut steps that can be done, and even improved, by some intelligent system. There has been a great increase in practices in this sense in the area of human health, which algorithms can perform some functions in favor of the agility and efficiency of the process.

To detect diabetic retinopathy, the patient should go to the doctor's office and take the background examination in which the doctor observes the image and identifies the disease. As this process is always the same, and some people do not have access to the exam, it is noted the possibility of automating it computationally with a system previously powered by a sick and non-sick imaging base.

This work can be improved by harvesting a larger number of images to cover all types of possible applications, and assisting to avoid overfitting, a practice that alone can enable the possibility of including an appropriate filter in the images, so that the system can better analyze the examples. The application of transfer learning proved to be profitable, and the testing of other pre-trained models can be a viable alternative for the continuation of this project, which will aim to increase the accuracy obtained by 82%. After improved accuracy, the application can be embarked on special devices that take the background examination of the eye, and in prototypes that take this examination to inhospitable regions, a process known as telemedicine, assisting people from these regions.

## 8.0 ATTACHMENTS

In this part, the attachments relevant to the work carried out will be added.

### 8.1 Attachment A – Specialist doctor report

The development of software capable of identifying signs of diabetic retinopathy is extremely important for the identification and treatment of this disease that is one of the main causes of blindness in Brazil and worldwide. With this system we were able to reach the population in the farthest places from large centers, greatly reducing the sequelae that diabetic retinopathy can cause.

Eduardo Henrique Marques Menezes
Ophthalmologist

## 8.2 Attachment B – Specialist doctor report

Eu, ______________________________ declaro que a proposta de TCC de Gilberto Luis De Conto Junior, com o titulo 'Detecção de retinopatia diabética por reconhecimento de imagem da retina', é um trabalho válido e viável.

Dr. Eduardo H. M. Menezes
Médico Oftalmologista
CRM/PR 26825

______________________________

Médico oftalmologista